\documentclass{ws-mpla}

\usepackage{slashed}

\begin{document}

\markboth{M.-C. Chen and J.R. Huang}
{TeV Scale Models of Neutrino Masses and Their Phenomenology}

\catchline{}{}{}{}{}

\title{TeV Scale Models of Neutrino Masses and Their Phenomenology
}

\author{\footnotesize Mu-Chun Chen and Jinrui Huang}

\address{Department of Physics and Astronomy, University of California\\
Irvine, California 92697-4575, U.S.A.\\
muchunc@uci.edu, jinruih@uci.edu}

\maketitle

\pub{Received (Day Month Year)}{Revised (Day Month Year)}

\begin{abstract}
We review various TeV scale models for neutrino masses utilizing different types of seesaw mechanisms, higher dimensional operators, expanded gauge symmetries, or low scale extra dimensions. In addition, we discuss the implications of these models for the collider experiments and low energy lepton flavor violation searches.

\end{abstract}

\ccode{PACS Nos.: 12.60.Cn; 14.60.Pq}

\section{Introduction}

In the conventional wisdom, the smallness of the neutrino masses is tied to the high scale of the new physics that generates neutrino masses. As this scale is very close to the grand unification (GUT) scale, it is generally not possible to directly test the model.  In view of the recent commissioning of the Large Hadron Collider (LHC), it is interesting to ask whether or not the neutrino mass generation can be due to new physics at the TeV scale so that one may probe the origin of neutrino mass generation at the collider experiments. 

In this review, we discuss various TeV scale mechanisms for neutrino mass generation. These include various seesaw mechanisms (Type-I, II, III) based on different UV completions of the Weinberg operator, radiative neutrino mass generation, MSSM with R-parity violation, models with expanded gauge symmetries including GUTs, TeV scale extra dimensions, or higher dimensional operator approach. Each of these scenarios provides distinct signatures that can be looked for at the collider experiments or in low energy lepton flavor violation searches.

The review is organized as follows. In Section \ref{sec:models}, we review various TeV scale mechanisms for neutrino mass generation.  In Section \ref{sec:leptg}, we comment on leptogenesis in the presence of a low seesaw scale.  Section \ref{sec:example} shows an explicit model based on TeV scale $U(1)^{\prime}$ family symmetry and its predictions for the LHC. This is followed by Section \ref{sec:sparticle} where the implications of the $U(1)^{\prime}$ family symmetry for the sparticle spectrum are discussed. Section \ref{sec:cond} concludes the review.

\section{TeV Scale Models for Neutrino Masses}\label{sec:models}

The mechanisms to lower the seesaw scale can be classified into the following classes according to their ultra-violet (UV) completion.

\subsection{Seesaw Mechanisms through Dimension Five Operators}

\begin{figure}[t!]
\centerline{\psfig{file=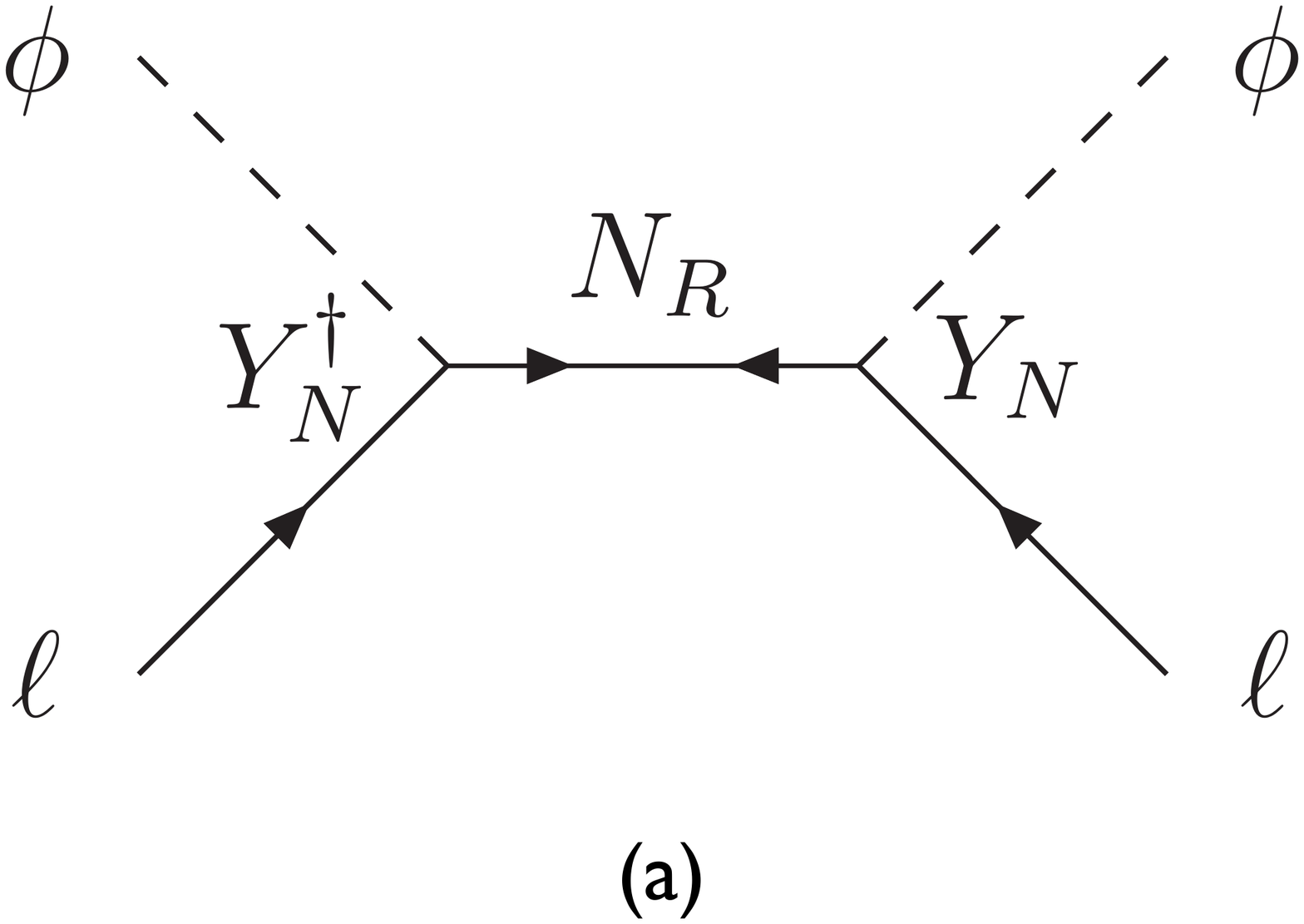,width=1.5in}\psfig{file=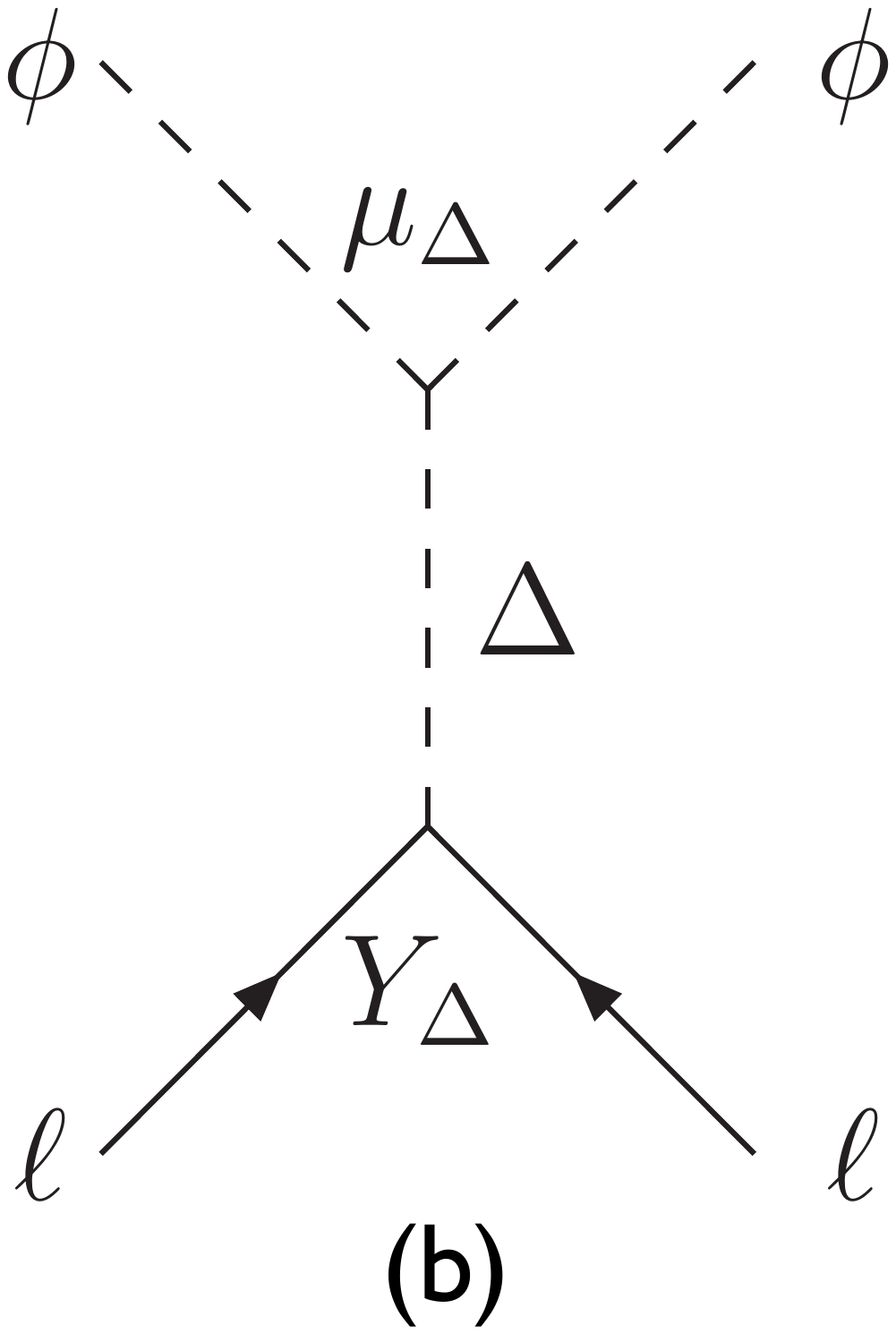,width=1.5in}\psfig{file=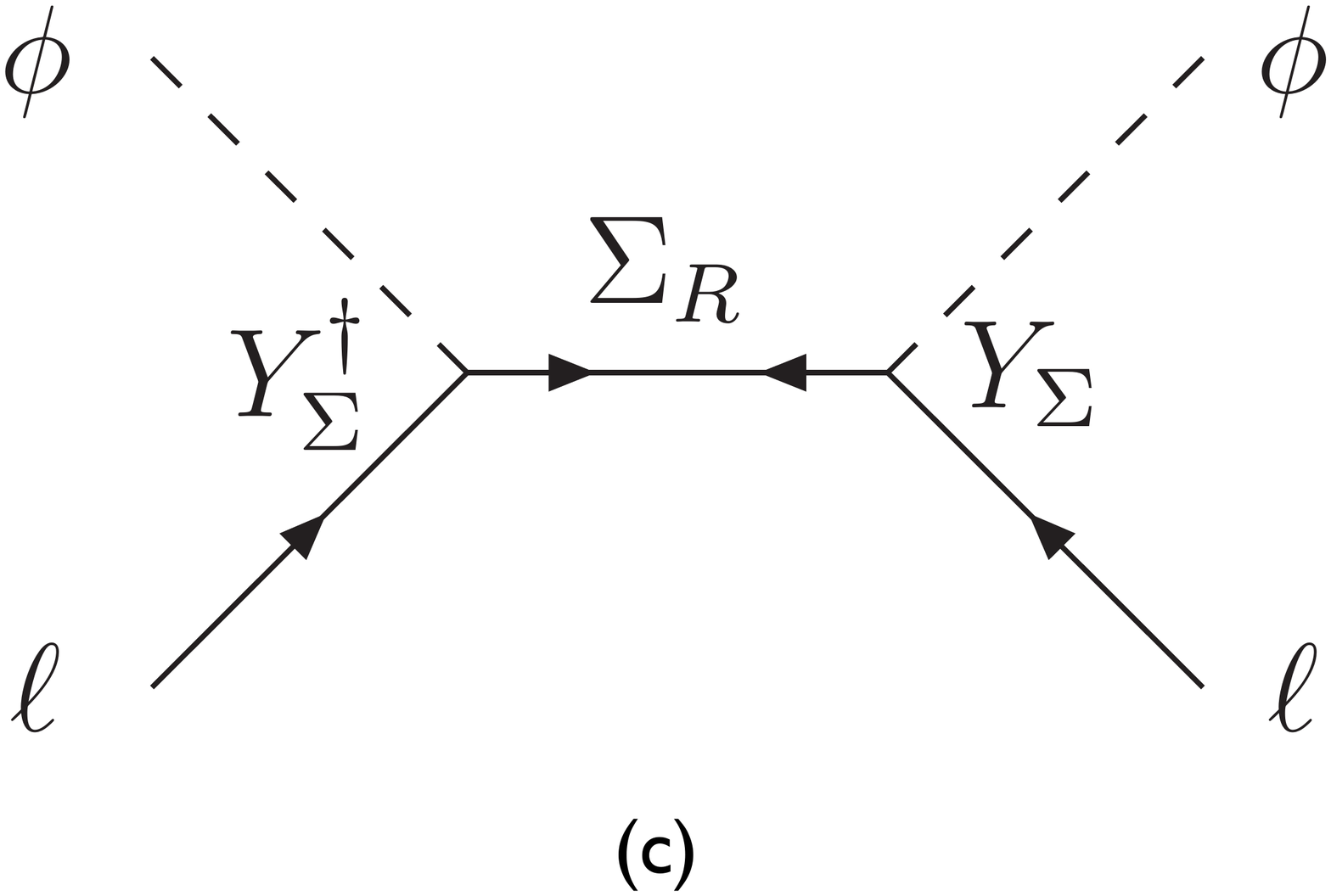,width=1.5in}}
\caption{The three types of UV completion for the dimension-5 Weinberg operator. (a) The type-I seesaw generates the operator through the exchange of the RH neutrino. (b) The type-II seesaw generates the operator through the exchange of a $SU(2)_{L}$ triplet Higgs. (c) The type-III seesaw generates the operator through the exchange of a $SU(2)_{L}$ triplet fermion.\protect\label{fig1}}
\end{figure}

If the effective neutrino masses are generated at dim-5, from naive dimensional analysis, the Yukawa coupling constants must be on the order of $10^{-6}$, to have the cutoff scale to be at the TeV scale. As a result, generically the seesaw sector is decoupled from the collider experiments, unless the production mechanism for the new particles that participate in the seesaw mechanism is independent of the small Yukawa couplings or small heavy-light mixing. This generally requires the existence of exotic states other than the right-handed neutrinos, or new interactions which right-handed neutrinos participate in. There are three UV completions of Weinberg operator, as shown in Fig.~\ref{fig1}, which correspond to Type-I, II, III. Different type of seesaw mechanisms may be distinguished by multi-lepton signals at the collider experiments\cite{delAguila:2008cj}. General discussions on collider searches for heavy Majorana neutrinos can be found in Ref.~\refcite{Atre:2009rg}.

\subsubsection{Type-I Seesaw}

In type-I seesaw mechanism, the field that UV completes the Weinberg operator is the right-handed neutrino. The only way to test seesaw mechanism is to produce the right-handed neutrinos. The effective light neutrino mass matrix, $m_{\mbox{\tiny eff}}$, is related to the RH Majorana mass matrix, $M_{\mbox{\tiny R}}$ and the Dirac mass matrix, $m_{\mbox{\tiny D}}$, as $m_{\mbox{\tiny eff}} = m_{\mbox{\tiny D}} M_{\mbox{\tiny R}}^{-1} m_{\mbox{\tiny D}}^{T}$. Given that the Yukawa coupling constants are small $\sim 10^{-6}$, the production of RH neutrinos through the Yukawa interactions is negligible. In the presence of the mixing between light and heavy neutrinos, it is also possible to produce the RH neutrinos through the gauge interactions. Nevertheless, the production is suppressed by the small mixing,
\begin{equation}
V = \frac{m_{\mbox{\tiny D}}}{M_{\mbox{\tiny R}}} \sim \frac{10^{-4} \; \mbox{GeV}}{100 \; \mbox{GeV}} = 10^{-6} \; ,
\end{equation}
while in order to have observable effects it generally requires\cite{del Aguila:2005pf}, 
\begin{equation}
V > 10^{-2} \; .
\end{equation} 
(Note that the constraints from the universality of the weak interactions and Z-width require $V < 10^{-1}$ can still be satisfied.)  
As a result, the physics responsible for neutrino mass generation is decoupled from the collider physics. 

To avoid having un-naturally small Yukawa couplings while retaining the RH neutrino mass scale in the TeV regime,  one may have the scenario in which at the leading order the effective neutrino mass matrix leads to three massless neutrinos. This can happen if and only if the Dirac neutrino mass matrix has rank one in the RH neutrino mass basis and that the contributions to the effective masses due to coupling to the three RH neutrinos  add up to zero\cite{Buchmuller:1991tu,Kersten:2007vk}. In this case, the Yukawa coupling constants can be arbitrarily large while having the RH neutrino mass scale at a TeV. Such cancellation may arise due to the presence of an underlying family symmetry, {\it e.g.} a discrete subgroup of the $U(1)_{L}$ symmetry\cite{Kersten:2007vk}, the $A_{4}$ symmetry\cite{Kersten:2007vk}, or the $S_{3}$ symmetry\cite{Adhikari:2010yt}. While at the leading order the seesaw mechanism predicts massless neutrinos, neutrinos acquire their masses through some other mechanisms. The collider signatures for lepton number violation include,
\begin{equation}
q \overline{q}^{\prime} \rightarrow \ell^{-}_{\alpha} \ell^{-}_{\beta} + \mbox{jets} \; .
\end{equation}
At the leading order, since neutrino masses are protected by a symmetry, the cross sections for these processes vanish. Neutrino masses are generated by some small perturbations. Due to the fact that these must be small perturbations to the leading order result,  the effects of lepton number violation are highly suppressed unless there is extreme fine-tuning\cite{Kersten:2007vk}.

\subsubsection{Type-II Seesaw}
\label{sec:type2}

In Type-II seesaw mechanism\cite{Mohapatra:1980yp}, the Weinberg operator is generated by the t-channel exchange of a $SU(2)_{L}$ triplet Higgs, $\Delta$, with hypercharge $Y=1$. The effective neutrino mass in this case is given by $Y \Delta LL$, where $Y$ is the Yukawa coupling constant. Upon the triplet acquiring a VEV,
\begin{equation}
v_{\Delta} = \frac{1}{\sqrt{2}} \mu v^{2} M_{\Delta}^{2} \; ,
\end{equation}
the neutrinos obtain effective masses given by $M_{\nu} = \sqrt{2} Y v_{\Delta}$.  
The parameter $\mu$ is the custodial symmetry breaking coupling in the scalar potential,  $\mu H\Delta H^{\dagger}$. Given that the VEV $v_{\Delta}$ must be small enough to give sufficient suppression to the effective neutrino masses, the precision EW constraint\cite{Chen:2003fm} from the $\rho$ parameter can be satisfied.

To have the mass scale of the triplet Higgs at a TeV requires $Y \mu \sim 10^{-12}$. If one allows the coupling constants to be as small as the electron Yukawa coupling, then $Y \sim \mu \sim \mathcal{O}(10^{-6})$. Upon the electroweak symmetry breaking taking place, there exist seven massive physical Higgs bosons: two neutral Higgses, $H_{1}, \; H_{2}$, one CP odd Higgs, $A$, two singlet charged Higgses, $H^{\pm}$, and two doubly charged Higgses, $H^{\pm\pm}$. 

The generic prediction of the model is the existence of the doubly charged Higgses, which couple only to the leptons, but not to the quarks. 
A unique signature of this class of model is that the doubly charged Higgses decay into same sign di-leptons (for a recent general discussion on the same sign dilepton signals at the collider experiments, see, Ref.~\refcite{Rentala:2011mr}),
\begin{equation}
\Delta^{\pm\pm} \rightarrow   \ell^{\pm} \ell^{\pm}, \quad (\ell = e, \; \mu, \; \tau) \; 
\end{equation}
which do not have any SM or MSSM backgrounds.  
As pointed out in Ref.~\refcite{Akeroyd:2005gt}, the doubly charged Higgses can be produced at the LHC via the Drell-Yan,
\begin{equation}
q\overline{q} \rightarrow \gamma^{\ast}, \; Z^{\ast} \rightarrow H^{++} H^{--}, \quad 
q\overline{q}^{\prime} \rightarrow W^{\ast} \rightarrow H^{\pm\pm} H^{\mp} \; .
\end{equation}
As the production of the triplet Higgs is through the gauge interactions, it is independent of the small light-heavy neutrino mixing and consequently can have unsuppressed production cross section, in contrast to the case of the Type-I seesaw. It has been shown that, for a triplet mass in the range of (200-1000) GeV, the cross section can be 0.1-100 fb. With 300 fb${}^{-1}$, a doubly charged Higgs, $\Delta^{++}$, with mass of 600 GeV can be discovered at the LHC. 

Phenomenology associated with the triplet Higgs at a linear collider has also been investigated\cite{Rodejohann:2010bv}. 

\subsubsection{Type-III Seesaw}

The Weinberg operator can also be UV completed by the mediation of a $SU(2)_{L}$ triplet fermion, $\Sigma = (\Sigma^{+}, \Sigma^{0}, \Sigma^{-})$, with zero hypercharge\cite{Foot:1988aq}.  The effective neutrino mass is $y_{\nu}^{2} v^{2} / \Lambda$, where $y_{\nu}$ is the Dirac Yukawa coupling of the triplet lepton to the SM lepton doublet and the Higgs and $\Lambda$ is the lepton number violation scale. To have $\Lambda \sim 1$ TeV, $y_{\nu}$ has a value $\sim 10^{-6}$.
   
Because the triplet lepton $\Sigma$ has weak gauge interactions, their production cross section is unsuppressed, contrary to the case of the Type-I seesaw. The signature with relatively high rate is\cite{Franceschini:2008pz}
\begin{equation}
pp \rightarrow \Sigma^{0} \Sigma^{+} \rightarrow \overline{\nu} W^{+} W^{\pm} \ell^{\mp} \rightarrow 4 \; \mbox{jets} 
+ \slashed{E}_{T} + \ell \; .
\end{equation}  
As the masses of $\Sigma^{\pm}$ and $\Sigma^{0}$ are on the order of  sub-TeV region, the displaced vertices from the primary production vertex in the $\Sigma^{0}$, $\Sigma^{\pm}$ decays can be visible\cite{Franceschini:2008pz}. The triplet lepton lifetime is related to the effective neutrino mass spectrum  
\begin{equation}
\tau \le 1 \; \mbox{mm} \times \biggl( \frac{0.05 \; \mbox{eV} }{ \sum_{i} m_{i} }\biggr)
\biggl( \frac{100 \; \mbox{GeV} }{ \Lambda } \biggr)^{2} \; .
\end{equation}
For the normal hierarchy case ($\sum_{i} m_{i} \simeq 0.05$ eV), this leads to $\tau \le 1 \; \mbox{mm}$ for $\Lambda \simeq 100$ GeV. (For other collider studies, see Ref.~\refcite{Li:2009mw}.) 
In addition, in the supersymmetric case, it has been pointed out that\cite{Chun:2009mh} the neutral component of the super-partner of the triplet lepton, $\tilde{\Sigma}^{0}$, can be a realization of the minimal dark matter\cite{Cirelli:2005uq}.  

Due to the mixing between the triplet lepton and the SM lepton doublets, tree level flavor changing neutral currents (FCNCs) are present in models with type-III  seesaw\cite{Abada:2007ux}. Constraints from LFV processes such as $\ell_{i} \rightarrow \ell_{j} + \gamma$, $\mu-e$ conversion, {\it etc.} have been investigated\cite{He:2009tf}. 

Type-III seesaw has been utilized in models with family symmetries, including a $\mu-\tau$ symmetry model\cite{Bandyopadhyay:2009xa} and a $A_{4}$ symmetry model\cite{Ahn:2011pq}. It can also naturally be incorporated in models with anomaly mediated SUSY breaking\cite{Mohapatra:2008wx}.

\subsection{Inverse Seesaw Mechanism}

In the so-called inverse seesaw mechanism\cite{Mohapatra:1986aw,Mohapatra:1986bd}, with the addition of an extra singlet $S$ for each generation besides a RH neutrino $\nu_{R}$, the following $9 \times 9$ neutrino mass matrix can be generated, in the basis of $(\nu_{L}, \; \nu_{R}, \; S)$, 
\begin{equation}
M_{\nu} = \left(
\begin{array}{ccc}
0 & M_{\mbox{\tiny D}} & 0 \\
M_{\mbox{\tiny D}}^{T} & 0 & M_{\mbox{\tiny NS}} \\
0 & M_{\mbox{\tiny NS}} & M_{\mbox{\tiny S}}
\end{array}\right) \; .
\end{equation} 
Here the Majorana mass term for $\nu_{R}$ is forbidden. It is possible to have a large Dirac mass, $M_{\mbox{\tiny D}}$, and TeV scale RH neutrino masses, if the following condition is satisfied,
\begin{equation}
M_{\mbox{\tiny S}} \ll M_{\mbox{\tiny D}} \ll M_{\mbox{\tiny NS}} \; .
\end{equation} 
The effective light neutrino mass matrix is given by, to the leading order,
\begin{equation}
M_{\mbox{\tiny eff}} \simeq (M_{\mbox{\tiny D}} M_{\mbox{\tiny NS}}^{-1}) M_{\mbox{\tiny S}} (M_{\mbox{\tiny D}} M_{\mbox{\tiny NS}}^{-1})^{T} \; .
\end{equation}
In other words, the smallness of the neutrino masses is due to the smallness of the lepton number violation coupling, $M_{\mbox{\tiny S}}$, which is  lower than the EW scale. Viable effective neutrino masses can be obtained with $M_{\mbox{\tiny NS}} \sim \mathcal{O}(1 \; \mbox{TeV})$, $M_{\mbox{\tiny D}} \sim \mathcal{O}(100 \; \mbox{GeV})$, and $M_{\mbox{\tiny S}} \sim \mathcal{O}(0.1 \; \mbox{keV})$.

In the inverse seesaw framework, sizable non-unitarity effects \cite{Malinsky:2009gw} and lepton flavor violation\cite{Ilakovac:1994kj} are expected. In addition, in a supersymmetric model of this type, a strong correlation is found between the lightest chargino decay widths and the widths of the lepton flavor violating charged lepton decays\cite{Hirsch:2009ra},
\begin{equation}
 \frac{\mbox{BR}(\tilde{\chi}_{1}^{\pm} \rightarrow \tilde{N}_{1+2} + \mu^{\pm})}{\mbox{BR}(\tilde{\chi}_{1}^{\pm} \rightarrow \tilde{N}_{1+2} + \tau^{\pm})} 
 \propto \frac{\mbox{BR}(\mu \rightarrow e + \gamma)}{\mbox{BR}(\tau \rightarrow e + \gamma)} \; .
\end{equation}
In both SUSY\cite{Deppisch:2004fa} cases and a non-SUSY\cite{Abdallah:2011ew} case with inverse seesaw, the branching fractions of the charged lepton flavor violating decays, $\ell_{i} \rightarrow \ell_{j} + \gamma$, are found to be enhanced. Implications for neutrinoless double beta decay have been investigated in Ref.~\refcite{Ibarra:2010xw}.

\subsection{Radiative Seesaw}

The smallness of the neutrino masses can also be explained if the neutrino masses are generated radiatively\cite{Zee:1985rj}. This is achieved in  Ref.~\refcite{Zee:1985id} at two-loops by having additional singly-charged $SU(2)_{L}$ singlet scalar fields and doubly-charged $SU(2)_{L}$ singlet scalar fields (Zee-Babu Model). With an additional $Z_{2}$ symmetry, it is also possible for the light neutrino masses to arise only at the higher loop levels with TeV scale RH neutrinos\cite{Krauss:2002px,Ma:2006km,Aoki:2008av}. Given that the new particles introduced in these TeV scale models differ model by model, the collider signatures\cite{Aoki:2010tf} are quite model dependent. It is to be noted that in the class of models with $Z_{2}$ symmetry, there is naturally a dark matter candidate\cite{Krauss:2002px,Ma:2006km,Aoki:2008av,Suematsu:2010gv}. The new particles involved in the loop may also be charged under the color $SU(3)$\cite{FileviezPerez:2010ch}. In this case, the production cross section can be enhanced.

Radiative neutrino mass generation described above can naturally be embeded into models with Coleman-Weinberg symmetry breaking\cite{Foot:2007ay}. The new TeV scale scalars required  achieve radiative EW symmetry breaking also contribute to the generation of neutrino masses.

\subsection{MSSM with R-Parity Violation}

Neutrino mass generation can also arise in models\cite{deCampos:2007bn} with R-parity violation, through the Bi-linear lepton number violating operators,
\begin{equation}
\mathcal{W}_{R} =  \epsilon_{i} \hat{L}_{i} \hat{H}_{u} \; ,
\end{equation}
where $\epsilon_{i}$ are coefficients of the operators of unit of mass.  
As the above operators are the only R-parity violating operator allowed in the model, proton decay is not induced. In a specific minimal realization\cite{DeCampos:2010yu} in MSSM with the Bi-linear lepton number violating operators, a correlation is found\cite{Porod:2000hv} between the atmospheric mixing angle and branching fractions of neutralino decays,
\begin{equation}
\tan^{2} \theta_{\mbox{\tiny atm}} \simeq \frac{\mbox{BR}(\tilde{\chi}_{1}^{0} \rightarrow \mu^{\pm} W^{\mp})}{\mbox{BR}(\tilde{\chi}_{1}^{0} \rightarrow \tau^{\pm} W^{\mp})} \; ,
\end{equation}
as the scale of  $\Delta m_{\mbox{\tiny atm}}^{2}$ is generated at tree level through the exchange of a weak scale neutralino. The scale of $\Delta m_{\odot}^{2}$ arises radiatively. At the LHC with 100 fb${}^{-1}$ at $14$ TeV, it is possible to probe a large fraction of the parameter space admitted by the neutrino oscillation data in this scenario.

\subsection{TeV Scale Extra Dimension}

Warped extra dimension is an alternative to supersymmetry as a solution to the gauge hierarchy problem, which requires the scale of the first Kaluza-Klein (KK) mode is on the order of a TeV. Due to the small overlap between the wave functions of the lepton doublets and the RH neutrinos, small neutrino masses of the Dirac type can naturally be generated\cite{Grossman:1999ra}. (Neutrinos of the Majorana type can also be accommodated, see Ref.~\refcite{Chen:2005mz}). By localizing different generations of SM fermions at different locations in the extra dimension as determined by the 5D bulk mass terms, the mass hierarchy can naturally arise\cite{Gherghetta:2000qt}. 

The non-universal 5D bulk mass terms lead to dangerously large flavor-changing neutral currents (FCNCs) at the tree level, leading to a stringent constraint on the first KK scale to be above $\mathcal{O}(10)$ TeV\cite{Kitano:2000wr}. In addition, to simultaneously obtain the mild neutrino mass hierarchy and two large and one small mixing leptonic mixing angles generally requires severe fine-tuning in the 5D Yukawa coupling constants. These problems are alleviated by imposing minimal flavor violation assumption in the quark\cite{Fitzpatrick:2007sa} and lepton\cite{Chen:2008qg} sectors.  These problems can also be avoided by imposing a bulk family symmetry based on $A_{4}$ in the lepton sector\cite{Csaki:2008qq}, or $T^{\prime}$ for both quarks and leptons\cite{Chen:2009gy}. 

In a particular setup\cite{Matsumoto:2010zg} with the RH neutrinos being the bulk fields while all SM fermions on the TeV brane, the tri-lepton final states are the most effective LHC signal,
\begin{equation}
pp \rightarrow \ell^{\pm} \ell^{\pm} \ell^{\mp} \nu (\overline{\nu}) \; .
\end{equation}
At the ILC with the center of mass energy of 500 GeV, the first KK state of the right-handed neutrinos may be produced through the channel
\begin{equation}
e^{+} e^{-} \rightarrow \nu N, \quad N \rightarrow W \ell \; ,
\end{equation}
while with the center of mass energy of 1 TeV, the second and third KK states may be produced\cite{Saito:2010xj}. 
The single lepton plus two-jet final states with large missing transverse energy provides a good channel to measure the masses and cross sections of the KK neutrinos. 

Neutrino mass generation in the split fermion scenario in extra dimension has been discussed in Ref.~\refcite{Frere:2003hn}. In Ref.~\refcite{Frere:2010ah}, in a 6D model in the vortex background on a sphere, a single fermion family in 6D can give rise to 3 zero modes, leading to a potential explaination of three families. In this setup, neutrinos are predicted to have an inverted pseudo-Dirac mass pattern and $U_{e3}$ is predicted to be $\sim 0.1$.

\subsection{Models with Expanded Gauge Symmetries}

In seesaw models with expanded gauge symmetries beyond those of the Standard Model, right-handed neutrinos can be charged under the new gauge symmetries and thus have enhanced production cross section compared to the case with only Standard Model gauge symmetries. 

\subsubsection{TeV Scale Left-right Symmetric Models}

The Type-II seesaw mechanism in the presence of a $SU(2)_{L}$ triplet Higgs as described in Sec.~\ref{sec:type2} can be naturally embedded into the left-right symmetry model with gauge symmetry $SU(2)_{L} \times SU(2)_{R} \times U(1)_{B-L}$. The $SU(2)_{R} \times U(1)_{B-L}$ symmetry is broken at the TeV scale, leading to a triplet Higgs, a $Z^{\prime}$ boson, as well as a $W^{\prime}$ boson (the $W_{R}$), all have a mass accessible to the collider experiments. (For constraints on the $Z^{\prime}$ and $W^{\prime}$ masses, see {\it e.g.} Ref.~\refcite{Chen:2008zz} and \refcite{Chen:2008zzx}). The connection to neutrinoless double beta decay has also been studied\cite{Tello:2010am}. A Type-I seesaw realization in such a TeV scale left-right symmetric model can be found in Ref.~\refcite{Chakrabortty:2010zk}. 

A TeV scale model with $U(1)_{B-L}$ gauge symmetry has also been investigated\cite{Huitu:2008gf}.

\subsubsection{GUT Models with TeV scale Exotic States}

It has been pointed out recently that\cite{Dev:2009aw} the unification of the three gauge coupling constants can still be retained even if some of the GUT exotic states have masses much lower than the GUT scale. In a specific model\cite{Dev:2009aw} based on SUSY SO(10), the intermediate left-right symmetry breaking scale is at the TeV regime, instead of the GUT scale. Neutrino masses are generated through the inverse seesaw mechanism, with RH neutrino masses on the order of a TeV.

\subsection{Higher Dimensional Operator Approach}

In the conventional seesaw mechanism, as the neutrino masses are generated by dim-5 effective operators, 
\begin{equation}
\frac{HHLL}{\Lambda} \; ,
\end{equation}
the cutoff scale of the new physics must be close to the GUT scale in order to sufficiently suppress the effective light neutrino masses. In the presence of some new symmetry, it is possible\cite{Carone:1996ny,Chen:2006bv,Chen:2006hn} to forbid operators with lower mass dimensionalities and neutrino masses are allowed only at high mass dimensionalities.  In this case, the operators generically have the suppression factor in terms of some power $p$ of the ratio of VEV of the scalar field, $\phi$, that breaks the new symmetry to the cutoff scale of the symmetry scale, $\left(\frac{\left< \phi \right>}{\Lambda}\right)^{p}$. If the dimensionality is high enough, the cutoff scale of the new physics, can be on the order of a TeV. In Section~\ref{sec:example} we show an explicit TeV scale model based on the higher dimensional approach, in the presence of an extra $U(1)^{\prime}$ symmetry.

In the presence of two Higgs doublets, a systematic classification of higher dimensional operators that are relevant for neutrino mass generation can be found in Ref.~\refcite{Bonnet:2009ej}.

\section{Low Seesaw Scale and Leptogenesis}\label{sec:leptg}

In standard leptogenesis, to generate sufficient amount of the lepton number asymmetry requires a heavy right-handed neutrino mass scale, generally on the order of $\sim 10^{9}$ GeV. (For a review on leptogenesis, see, {\it e.g.} Ref.~\refcite{Chen:2007fv}.) In this high scale scienario, leptogenesis can be tested only through archaeological evidences\cite{Frere:2008ct}. 
With the seesaw scale lowered to the TeV regime, one immediate question is whether or not leptogenesis can arise.  While it is difficult\cite{Hambye:2001eu} to implement the standard leptogenesis in the presence of a low lepton number violation scale, there are alternative leptogenesis mechanisms that can be implemented at the low scale. 

For example, if two right-handed neutrinos have near degenerate masses, the lepton number asymmetry is enhanced due to the resonant effects in the self-energy diagram. This is the so-called resonant leptogenesis\cite{Pilaftsis:1997jf}. It has been shown\cite{Pilaftsis:2003gt} that due to the resonance enhancement, sufficient asymmetry can be generated even if the right-handed neutrinos have masses on the order of a TeV. Consequently, there are possible signatures at the collider experiments\cite{Chakrabortty:2011zz}.

In addition, it has been pointed out\cite{Blanchet:2010kw} that in the inverse seesaw scenario, the lepton number violating wash-out processes vanish due to the low lepton number violation scale.    

Soft leptogenesis\cite{D'Ambrosio:2003wy} can also be incorporated in models with TeV scale seesaw. A realization in a model with inverse seesaw can be found in Ref.~\refcite{Garayoa:2006xs}.

\section{An Explicit Example with TeV Scale $U(1)^{\prime}$ Family Symmetry}
\label{sec:example}

In the sections above, we review various mechanisms to lower the seesaw scale down to the TeV region. While these mechanisms can give rise to the suppression required in the neutrino masses, in order to explain the mixing pattern, additional new physics, such as a family symmetry, is required. Family symmetries that have been utilized include continuous groups such as $SU(2)$\cite{Chen:2000fp}, or finite groups, such as $A_{4}$\cite{Chen:2009um}, or $T^{\prime}$\cite{Chen:2007afa,Chen:2009gf}, which afford the possibility of a geometrical origin of CP violation\cite{Chen:2009gf}. Different models may be distinguished by precise measurements of the mixing parameters\cite{Albright:2006cw} or LFV processes\cite{Albright:2008ke}. 

Here we present a supersymmetric model\cite{Chen:2009fx} with an extra $U(1)$ symmetry with non-universal $U(1)^{\prime}$ charges for different generations of SM fermions. Due to the $U(1)^{\prime}$ symmetry, neutrino masses can arise only through higher dimensional operators\cite{Chen:2006hn}. The $U(1)^{\prime}$ also plays  the role of a family symmetry  giving rise to realistic mass hierarchy and mixing angles among the SM fermions, including the neutrinos, through the Froggatt-Nielsen (FN) mechanism\cite{ref:frogNiel}. (For review, see {\it e.g.} Ref.~\refcite{Chen:2010bt}.) The $U(1)^{\prime}$ charge assignment naturally suppresses the $\mu$ term, and it forbids at the tree level baryon number and lepton number violating operators that could lead to proton decay.

Models with an extra $U(1)^{\prime}$ symmetry at the TeV scale are severely constrained by the flavor-changing-neutral-current (FCNC) processes and by the electroweak precision measurements. The FCNC constraints can be satisfied by attributing the flavor mixing to the up-type quark and neutrino sectors, while having flavor diagonal down-type quark and charged lepton sectors, given that the down-type quark and charged lepton sectors are most stringently constrained by the measurement of $K^{0}-\overline{K}^{0}$ and $B^{0}-\overline{B}^{0}$ mixing in the quark sector, and in the lepton sector from the non-observation of $\mu-e$ conversion as well as the LFV charged lepton decays. 

\subsection{The Model}

In MSSM with three right-handed neutrinos, the superpotential for the Yukawa sector and Higgs sector that gives masses to all SM fermions and Higgs fields is given as follows,
\begin{eqnarray}
\label{eqn:sPoten}
W  & = &  Y_uH_uQu^c + Y_dH_dQd^c + Y_eH_dLe^c  \; , \\
& & \quad + Y_{\nu}H_uL\nu^{c} 
+ Y_{LL}LLH_uH_u + Y_{\nu\nu}\nu^c\nu^c  \; . \nonumber
\end{eqnarray}
In the presence of the $U(1)^{\prime}$ symmetry, the Yukawa matrices in the superpotential shown above are the effective Yukawa couplings generated through higher dimensional operators {\it \`a la} the Froggatt-Nielsen mechanism. As a result, they can be written as powers of the ratio of the flavon fields, $ \Phi $ and $ \Phi^{\prime}$, that breaks the $U(1)^{\prime}$ symmetry, to the cutoff scale of the $U(1)^{\prime}$ symmetry, $\Lambda$,
\begin{equation}
\label{eqn:genYukawa1}
Y_{ij} \sim \biggl( y_{ij} \frac{  \Phi }{\Lambda} \biggr)^{3|q_i+q_j+q_H|} \; .
\end{equation} 
The chiral superfield $\Phi$ is a SM gauge singlet whose $U(1)^{\prime}$ charge is normalized to $-1/3$. The parameters $y_{ij}$ and $\mu_{ud}$ are coupling constants of order $\mathcal{O}(1)$; $q_i$ and $q_j$  are the $U(1)^{\prime}$ charges of the chiral superfields of the $i$-th and $j$-th generations of quarks and leptons, and $q_H$ (which can be $q_{H_u}$ or $q_{H_d}$) denotes the $U(1)^{\prime}$ charges of the up- and down-type Higgses. Note that if $q_{i}+q_{j}+q_{H} < 0$ or $q_{H_{u}} + q_{H_{d}} < 1/3$, then instead of the $\Phi$ field, the field $\Phi^{\prime}$ whose $U(1)^{\prime}$ charge is $1/3$ is used, so that the holomorphism of the superpotential is retained. The terms with non-integer $3|q_i+q_j+q_H|$ and $3|q_{H_u}+q_{H_d}|$ are not allowed in the superpotential given that the number of the flavon fields must be an integer. This thus naturally gives rise to texture-zeros in the Yukawa matrices. 

Once the scalar component $\phi$ ($\phi^{\prime}$) of the flavon superfield $\Phi$ ($\Phi^{\prime}$) acquires a vacuum expectation value (VEV), the $U(1)^{\prime}$ symmetry is broken. Upon the breaking of the $U(1)^{\prime}$ and electroweak symmetry, the effective Yukawa couplings then become,  
\begin{equation}
\label{eqn:genYukLam}
Y_{ij}^{eff} \sim \left(y_{ij}^3 \lambda \right)^{|q_i+q_j+q_H|} \; ,
\end{equation}
where $\lambda \equiv \left( \left< \phi \right> / \Lambda \right)^3$ or $\lambda \equiv \left( \left< \phi^{\prime} \right> / \Lambda \right)^3$.  
The $U(1)^{\prime}$ charges thus determine the form of the effective Yukawa matrices: 
Take the up-type quark Yukawa matrix as an example, which is, 
\begin{eqnarray}
Y_u & \sim & \left(\begin{array}{ccc} (\lambda)^{|q_{Q_1}+q_{u_1}+q_{H_u}|} & (\lambda)^{|q_{Q_1}+q_{u_2}+q_{H_u}|} & (\lambda)^{|q_{Q_1}+q_{u_3}+q_{H_u}|} \\ (\lambda)^{|q_{Q_2}+q_{u_1}+q_{H_u}|} & (\lambda)^{|q_{Q_2}+q_{u_2}+q_{H_u}|} & (\lambda)^{|q_{Q_2}+q_{u_3}+q_{H_u}|} \\ (\lambda)^{|q_{Q_3}+q_{u_1}+q_{H_u}|} & (\lambda)^{|q_{Q_3}+q_{u_2}+q_{H_u}|} & (\lambda)^{|q_{Q_3}+q_{u_3}+q_{H_u}|} \end{array} \right) \;.
\end{eqnarray}
In the neutrino sector, the Dirac, left-handed Majorana, as well as the right-handed Majorana mass matrices are given in terms of the $U(1)^{\prime}$ charges, respectively, as,
\begin{eqnarray}
\label{eqn:neutDiracYukawa}
Y_{\nu} & \sim & \left(\begin{array}{lll} (\lambda)^{|q_{f_1}+q_{N_1}+q_{H_u}|} & (\lambda)^{|q_{f_1}+q_{N_2}+q_{H_u}|} & (\lambda)^{|q_{f_1}+q_{N_3}+q_{H_u}|} \\ (\lambda)^{|q_{f_2}+q_{N_1}+q_{H_u}|} & (\lambda)^{|q_{f_2}+q_{N_2}+q_{H_u}|} & (\lambda)^{|q_{f_2}+q_{N_3}+q_{H_u}|} \\ (\lambda)^{|q_{f_3}+q_{N_1}+q_{H_u}|} & (\lambda)^{|q_{f_3}+q_{N_2}+q_{H_u}|} & (\lambda)^{|q_{f_3}+q_{N_3}+q_{H_u}|} \end{array} \right) \; ,
\\
\label{eqn:neutMajLeYukawa}
Y_{LL} & \sim & \left(\begin{array}{lll} (\lambda)^{|2q_{f_1}+2q_{H_u}|} & (\lambda)^{|q_{f_1}+q_{f_2}+2q_{H_u}|} & (\lambda)^{|q_{f_1}+q_{f_3}+2q_{H_u}|} \\ (\lambda)^{|q_{f_2}+q_{f_1}+2q_{H_u}|} & (\lambda)^{|2q_{f_2}+2q_{H_u}|} & (\lambda)^{|q_{f_2}+q_{f_3}+2q_{H_u}|} \\ (\lambda)^{|q_{f_3}+q_{f_1}+2q_{H_u}|} & (\lambda)^{|q_{f_3}+q_{f_2}+2q_{H_u}|} & (\lambda)^{|2q_{f_3}+2q_{H_u}|} \end{array} \right) 
\; , \\
\label{eqn:neutMajRiYukawa}
Y_{\nu\nu} & \sim & \left(\begin{array}{lll} (\lambda)^{|2q_{N_1}|} & (\lambda)^{|q_{N_1}+q_{N_2}|} & (\lambda)^{|q_{N_1}+q_{N_3}|} \\ (\lambda)^{|q_{N_2}+q_{N_1}|} & (\lambda)^{|2q_{N_2}|} & (\lambda)^{|q_{N_2}+q_{N_3}|} \\ (\lambda)^{|q_{N_3}+q_{N_1}|} & (\lambda)^{|q_{N_3}+q_{N_2}|} & (\lambda)^{|2q_{N_3}|} \end{array} \right) M_R \; ,
\end{eqnarray}
where $q_{Q_{i}}$, $q_{u_{i}}$, $q_{d_{i}}$, $q_{L_{i}}$, $q_{e_{i}}$, and $q_{N_{i}}$ denote, respectively, the charges of the quark doublet, iso-singlet up-type quark, iso-singlet down-type quark, lepton doublet, iso-singlet charged lepton, and right-handed neutrino of the $i$-th generation. With the experimental constraints on the fermion masses and mixing angles, the number of free parameters in the model is further reduced.

\subsection{Anomaly Cancellation}

Anomalous $U(1)^{\prime}$ family symmetry has been utilized extensively in flavor model building. It could be due to the fact that the $[U(1)^{\prime}]^{3}$ anomaly cancellation condition is difficult to solve (the integer solutions correspond to Fermat's last theorem.) In addition, it was stated in Ref.~\refcite{Ibanez:1994ig} that to give realistic fermion masses and mixing pattern, the $U(1)^{\prime}$ symmetry must be anomalous. Nevertheless, counter examples to this statement have been found, demonstrating that a non-anomalous $U(1)^{\prime}$ symmetry can be a viable family symmetry in both GUT\cite{Chen:2008tc} and non-GUT\cite{Chen:2006hn,Chen:2009fx,Chen:2010tf} models. In this case, constraints on the matter field charges are considerably more stringent compared to the case with anomalous $U(1)^{\prime}$. 

In the presence of the $U(1)^{\prime}$ symmetry, there are six additional anomaly cancellation conditions. All Higgs super-fields in the model are  assumed to appear in conjugate pairs and therefore do not contribute to the gauge anomalies. 
\begin{eqnarray}
\label{eqn:su3u1}
[SU(3)]^{2} U(1)^{\prime}_{F} & : &   \sum_{i} \left[ 2q_{Q_i} - (-q_{u_i}) - (-q_{d_i}) \right]  = 0 \; , 
\\
\label{eqn:su2u1}
[SU(2)_{L}]^{2} U(1)^{\prime}_{F} & : & \sum_{i} \left[ q_{L_i} + 3q_{Q_i}\right] = 0  \; , 
\\
\left[U(1)_{Y}\right]^{2} U(1)^{\prime}_{F} & : & 
\sum_{i} \biggl[ 2 \times 3 \times \biggl( \frac{1}{6} \biggr)^2 q_{Q_i} - 3 \times \biggl( \frac{2}{3} \biggr)^2 (-q_{u_i} ) \; 
 \label{eqn:u1y2u1}  \\
& &  - 3 \times \biggl( -\frac{1}{3}\biggr)^2 (-q_{d_i})
 + 2 \times \biggl(-\frac{1}{2}\biggr)^2 q_{L_i} - (-1)^2 (-q_{e_i}) \biggr] = 0 \; , \nonumber 
\end{eqnarray}
\begin{eqnarray}
\left[U(1)_{F}^{\prime}\right]^{2} U(1)_{Y} & : & 
\displaystyle \sum_{i} \biggl[ 2 \times 3 \times \biggl( \frac{1}{6} \biggr) q_{Q_i}^2 - 3 \times \biggl( \frac{2}{3} \biggr) \times (-q_{u_i})^2 
\; \label{eqn:u1yu12}\\
&&  - 3 \times \biggl(-\frac{1}{3} \biggr) (-q_{d_i})^2 
+ 2 \times \biggl(-\frac{1}{2}\biggr)(q_{L_i})^2 - (-1)(-q_{e_i})^2 \biggr] = 0 \; , 
\nonumber \\
\label{eqn:u1grav}
U(1)^{\prime}_{F}-\mbox{gravity} & : & 
\displaystyle \sum_{i} \left[ 6q_{Q_i} + 3q_{u_i} + 3q_{d_i} + 2q_{L_i} + q_{e_i} + q_{N_i}\right] = 0 \; , 
\\
\label{eqn:u13}
[U(1)^{\prime}_{F}]^{3} & : &  \hspace{-0.05in}
\sum_{i} \biggl[ 3 \bigl( 2 (q_{Q_i})^3 - (-q_{u_i})^3 - (-q_{d_i})^3\bigr) + 2(q_{L_i})^3 \; \\
&& - (-q_{e_i})^3 - (-q_{N_i})^3 \biggr] = 0\;. \nonumber
\end{eqnarray}
In order to find the solutions to the anomaly cancellation conditions, it is convenient to parametrize the  $U(1)^{\prime}$ charges in terms of the charge splitting\cite{Chen:2006hn,Chen:2009fx}. With this parametrization, all anomaly cancellation conditions are satisfied except for the $[U(1)^{\prime}]^{2} U(1)_{Y}$ and the $[U(1)^{\prime}]^{3}$ conditions. 

The anomaly cancellation conditions give rise to stringent constraints on the charges of the three generations of matter fields\cite{Chen:2009fx,Chen:2010tf}. They lead to highly predictive models for the flavor structure and mass hierarchy with a small number of free parameters.  

\subsection{Collider Phenomenology}

The $Z^{\prime}$ gauge boson associated with the breaking of the $U(1)^{\prime}$ symmetry is one of the exotic particles that are likely to be discovered at the early stages of the LHC. In addition, due to the generation-dependent nature of the $U(1)^{\prime}$ charges, this model possesses various flavor violating processes that can distinguish it from flavor-blind models.   

We provide a sample set of $U(1)^{\prime}$ charges in Table~\ref{tbl:sChargeZ} that satisfy all anomaly cancellation conditions and constraints on neutrino masses and mixing angles. As the $Z^{\prime}$ decay properties are dictated by the same set of $U(1)^{\prime}$ charges, the model can be tested by measuring these decay properties as demonstrated in the following.

\begin{table}[t!]
\tbl{The $U(1)_\nu$ charges of SM fermions, Higgs field H and scalar field $\phi$ in the specific case of $a = 13/3$, $b = -5/3$, and $g_{z^{\prime}} = 0.1$.}
{\begin{tabular}{c|c|c|c} \hline\
Field & $q_L, u_R, d_R$ & $\ell_{L_{1}}, e_{R_{1}}$ & $\ell_{L_{2,3}}, e_{R_{2,3}}$ \\ \hline
$U(1)_\nu$ charge & $z_q = -\frac{43}{72}$ & $z_{\ell_{1}} = -\frac{55}{8}$ & $z_{\ell_{2}}  = z_{\ell_{3}} = \frac{49}{8}$ \\ \hline \hline
Field & $\nu_{R_{1}}$ & $\nu_{R_{2,3}}$ & $H$, $\phi$\\ \hline
$U(1)_\nu$ charge & $z_{\nu_{1}} = \frac{41}{8}$ & $z_{\nu_{2}} = z_{\nu_{3}} = \frac{1}{8}$ & $z_H = 0$, $z_{\phi} = 1$  \\ \hline
\end{tabular} \label{tbl:sChargeZ}}
\end{table}

\subsubsection{The $Z^{\prime}$ Discovery} 

The $Z^{\prime}$ may be discovered by detecting excess signals from backgrounds near its resonance in the dilepton invariant mass distribution. The leading order cross section of the exclusive $Z^{\prime}$ production, $PP(q\bar{q}) \rightarrow Z^{\prime} \rightarrow e^{+}e^{-}/\mu^{+}\mu^{-}$, 
is given by,
\begin{eqnarray}
\label{eqn:xSec2}
\sigma(PP(q\bar{q}) \rightarrow Z^{\prime} \rightarrow \ell^{+} \ell^{-}) =  \frac{g_{z^{\prime}}^4z_q^2(z_{\ell_{m}})^2}{18\pi} \hspace{2in} \;  \\
\times \sum_{q=u,d,c,s,b} \int_0^1 \!\! \int_0^1 \,dx_1 \,dx_2 
f_q^{P}(x_1, M_{Z^{\prime}}^2)f_{\bar{q}}^P(x_2,M_{Z^{\prime}}^2) \frac{\hat{s}}{(\hat{s}-M_{Z^{\prime}}^2)^2+(M_{Z^{\prime}} \Gamma_{Z^{\prime}})^2} \; , \nonumber
\end{eqnarray}
where $f_q^{P}(x_1, Q^2)$ and $f_{\bar{q}}^{P}(x_2, Q^2)$ are parton distribution functions for the protons with $\hat{s} = x_1 x_2 s$.
Since the $Z^{\prime}$ decay width is narrow, the interference term between the $Z^{\prime}$ and the SM gauge bosons is neglected in the above equation. 

The main backgrounds which can mimic the signal events $Z^{\prime} \rightarrow e^{+}e^{-}$ and $Z^{\prime} \rightarrow \mu^{+}\mu^{-}$, can be categorized into two types, the reducible and irreducible backgrounds. The dominant QCD backgrounds, including inclusive jets, $W + $jets, $W+\gamma$, $Z + $jets, $Z+\gamma$, $\gamma + $jets, and $\gamma+\gamma$, can be avoided by applying various selection cuts (see Ref.~\refcite{ref:ATLASNote}.)  After these selection cuts, the dominant backgrounds are the irreducible backgrounds, which mainly come from the SM Drell-Yan processes. Other processes like decay products from $WW$, $WZ$, $ZZ$ and $t \bar{t}$ etc., can be ignored since their cross sections are very small. Hence, in the results shown below, $Z^{\prime} \rightarrow e^{+}e^{-}$ and $Z^{\prime} \rightarrow \mu^{+} \mu^{-}$ and the SM Drell-Yan are the only processes included in the PYTHIA simulations. 

\begin{figure}[t!]
{\center
\includegraphics[scale=0.8, angle = 90, width = 80mm, height = 50mm]{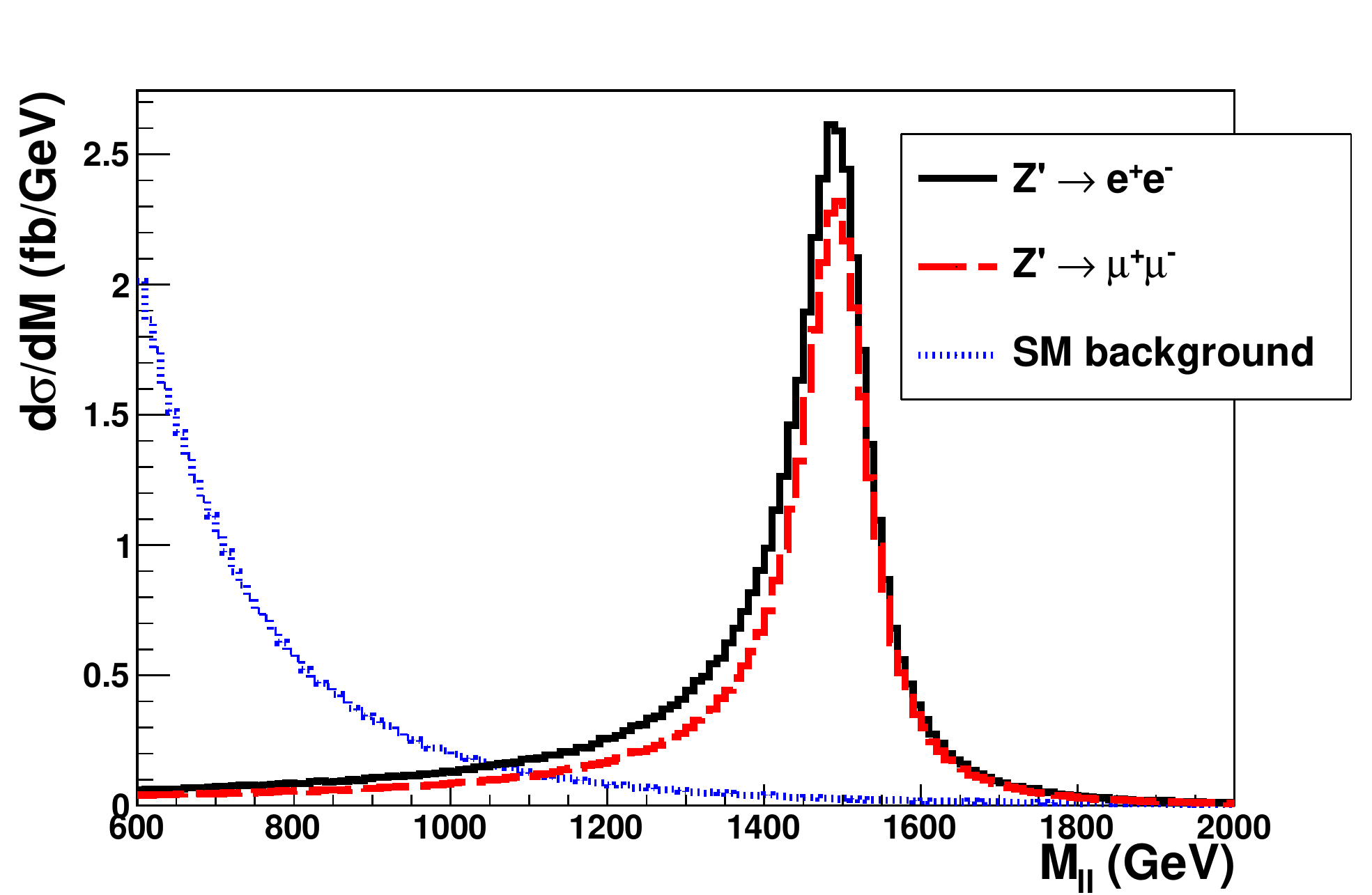}
    \caption{Dilepton invariant mass distributions $M_{Z^{\prime}} = 1.5$ TeV at $\sqrt{s} = 14$ TeV. The black solid line represents the $Z^{\prime} \rightarrow e^{+}e^{-}$ channel, while the red dashed line is the $Z^{\prime} \rightarrow \mu^{+}\mu^{-}$ channel. The blue dotted line is the SM backgrounds for both channels, which are almost identical for both channels at the tree level since $Z$ and $\gamma$ couple to SM fermions universally.\protect\label{fig:invM}}}    
\end{figure}

The dilepton invariant mass distributions in the $e^{+}e^{-}$ and $\mu^{+}\mu^{-}$ channels for $M_{Z^{\prime}} = 1.5$ TeV are shown in Fig~\ref{fig:invM}. The SM backgrounds, which are almost identical for both channels at the tree level since the gauge couplings of $Z$ and $\gamma$ to the SM fermions are universal, are highly suppressed in the $Z^{\prime}$ resonance region, allowing a clear distinction between the signals from the backgrounds.

A $5\sigma$ discovery of the $Z^{\prime}$ at the LHC entails $S/\sqrt{B} \equiv \sigma_{S}*L/\sqrt{\sigma_{B}*L}$, with $S$ being the number of $Z^{\prime}$ signal events that satisfies $|M_{\ell \ell} - M_{Z^{\prime}}| < 2\Gamma_{Z^{\prime}}$, $B$ being the number of SM Drell-Yan background events, $\sigma_{S}$ and $\sigma_{B}$ respectively being the total cross sections of the signal and background events, and $L$ being the integrated luminosity. With $100$ fb$^{-1}$ of data, the LHC can discover a $Z^{\prime}$ with a mass up to 4.5 TeV at the center of mass energy $\sqrt{s} = 14$ TeV.

\subsubsection{Testing the Flavor Dependence}

While there is a great potential for the $Z^{\prime}$ discovery at the LHC during its early low luminosity running, to establish the flavorful nature of the $Z^{\prime}$, after it has been discovered in the dilepton channels, requires a significant larger amount of data. 

\begin{figure}[t!]
{\center
\includegraphics[scale=0.8, angle = 90, width = 80mm, height = 50mm]{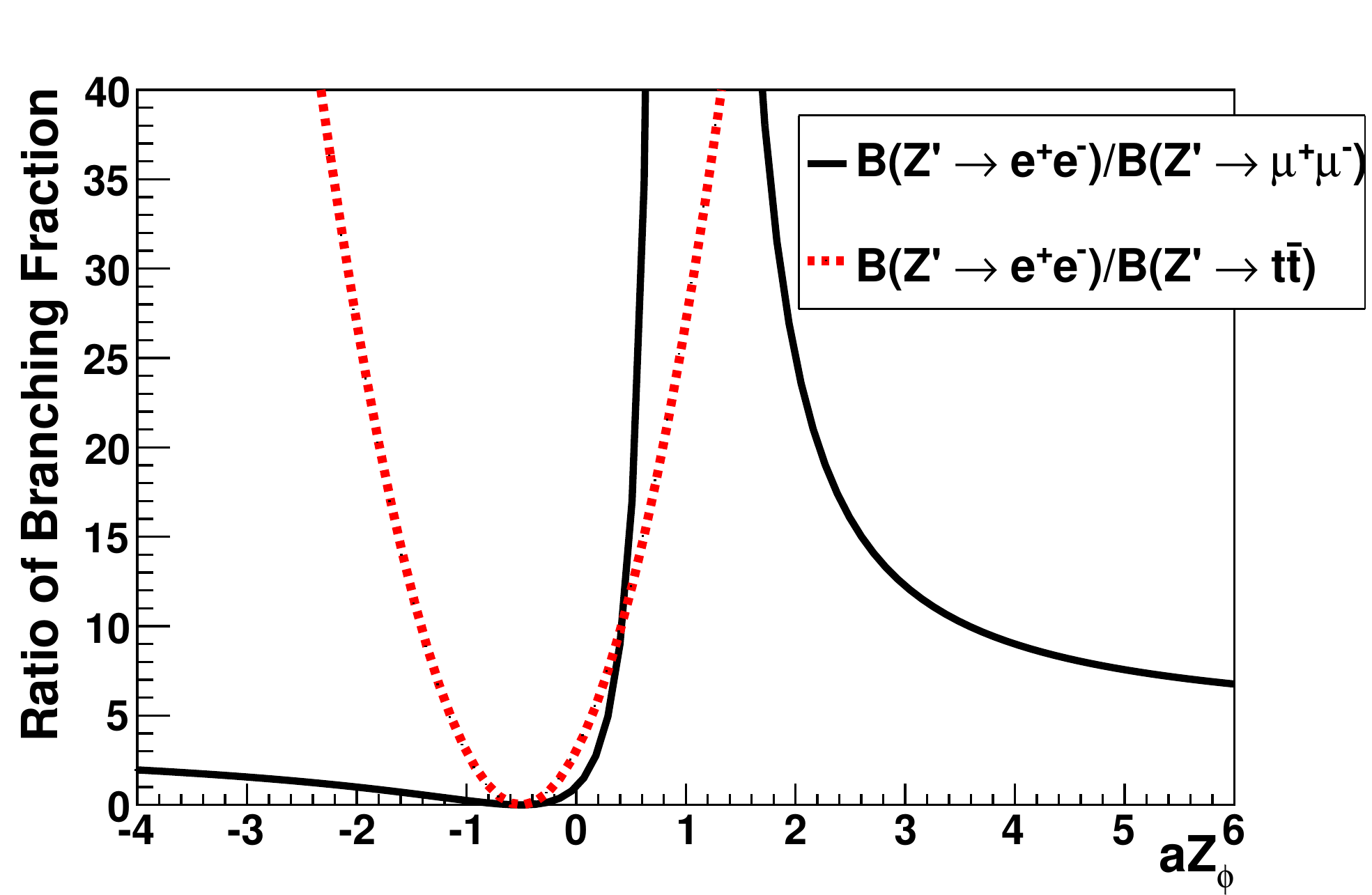}
    \caption{Ratios of branching fractions ($R_{e\mu}$ and $R_{et}$) as a function of variable $Z_{\phi} = \frac{-3(a+b)}{a^{2}+ab+b^{2}}$.\protect\label{fig:BRRatio}}}    
\end{figure}

The flavorful nature of the $Z^{\prime}$ can be established in the model by measuring the ratio of the decay branching fractions and forward-backward asymmetry. In the $U(1)_{\nu}$ model, the electron and muon are allowed to have different $U(1)_{\nu}$ charges, the ratio of the decay branching fractions, 
$R_{e\mu} = B(Z^{\prime} \rightarrow e^{+}e^{-})/B(Z^{\prime} \rightarrow \mu^{+}\mu^{-})  \; , $
can in general differ from $1$. The ratio $R_{e\mu}$ is given as,
\begin{equation}
R_{e\mu} = \frac{B(Z^{\prime} \rightarrow e^{+}e^{-})}{B(Z^{\prime} \rightarrow \mu^{+} \mu^{-})} 
= \biggl(\frac{z_{\ell_1}}{z_{\ell_2}}\biggr)^2 = \biggl( \frac{1+2aZ_{\phi}}{1-aZ_{\phi}} \biggr)^2  \; .
\end{equation}
If the $Z^{\prime}$ is also discovered in the $t\overline{t}$ channel, by measuring both $R_{e\mu}$ and the ratio of the decay branching fractions of 
$e^{+}e^{-}$ to $t\overline{t}$ channels, 
\begin{equation}
R_{et} = \frac{B(Z^{\prime} \rightarrow e^{+}e^{-})}{B(Z^{\prime}  \rightarrow t \bar{t})} = 3(1+2aZ_{\phi})^2 \; , 
\end{equation}
and we can uniquely determine the parameter $aZ_{\phi} = -3(a+b)/(a^{2}+ab+b^{2})$, as demonstrated in Fig.~\ref{fig:BRRatio} and further determine the neutrino mixing angle. 

The bench mark point, $z_{\ell_{1}} = -55/8$ and $z_{\ell_{2}} = 49/8$, predicts a ratio of the branching fractions of 
$R_{e\mu} = (55/49)^2 \approx 1.26$. To study the integrated luminosity required for distinguishing the $e^{+}e^{-}$ and $\mu^{+}\mu^{-}$ channels at $5\sigma$ using the counting method, we define the following variable, 
\begin{equation}
\frac{D}{\sqrt{B}} = \frac{\sigma_{ee}\ast L - \sigma_{\mu\mu} \ast L}{\sqrt{\sigma_{\mu\mu} \ast L}} \; ,
\end{equation}
where $\sigma_{ee}$ and $\sigma_{\mu\mu}$ are the cross sections of the $e$ and $\mu$ events. The parameter $D$ thus gives the difference in the numbers of events between $Z^{\prime} \rightarrow e^{+}e^{-}$ and $Z^{\prime} \rightarrow \mu^{+}\mu^{-}$.  If $D/\sqrt{B} > 5$, the difference between the dielectron and dimuon events is above $5 \sigma$ significance. At $\sqrt{s} = 14$ TeV  with $500$ fb$^{-1}$ of data, a statistically significant distinction between the branching fractions for $e$ and $\mu$ channels can be obtained up to $M_{Z^{\prime}} = 3$ TeV. 

Additionally, forward-backward asymmetry $A_{\mbox{\tiny FB}}$ can also be used to distinguish various flavor $U(1)^{\prime}$ models. It is defined as,
\begin{equation}
A_{FB} \equiv \frac{\sigma_{F} - \sigma_{B}}{\sigma_{F} + \sigma_{B}} \; ,
\end{equation}
where $\sigma_{F, B}$ are the total cross sections of the forward and backward events, respectively.  By using the forward backward asymmetry distributions, a clear distinction between the dielectron and dimuon channels can be obtained in the low invariant mass and low transverse momentum regions with 500 fb$^{-1}$ data. This is shown in Fig. ~\ref{fig:AFB1} and ~\ref{fig:AFB2}.

\begin{figure}[t!]
\label{fig:AFB1}
{\center
\includegraphics[scale=0.8, angle = 90, width = 80mm, height = 50mm]{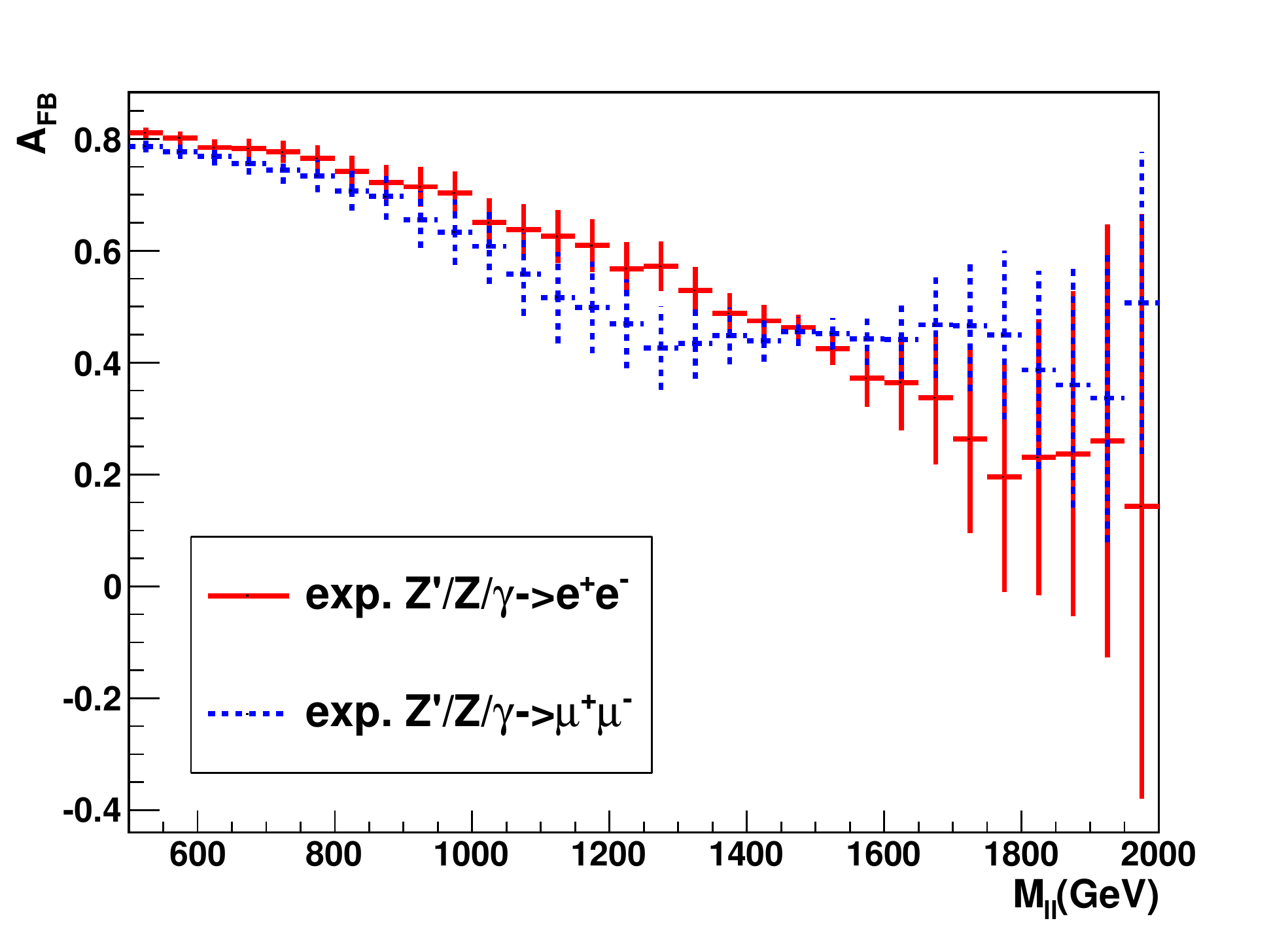}
    \caption{The forward backward asymmetry distribution as a function of the dilepton invariant mass for $M_{Z^{\prime}} = 1.5$ TeV in the case of $\sqrt{s} = 14$ TeV with $500$ fb$^{-1}$ of data. The red lines represent the electron channel and the dotted blue lines denote the muon channel. \protect\label{fig:AFB1}}}
\end{figure}

\begin{figure}[t!]
\label{fig:AFB2}
{\center
\includegraphics[scale=0.8, angle = 90, width = 80mm, height = 50mm]{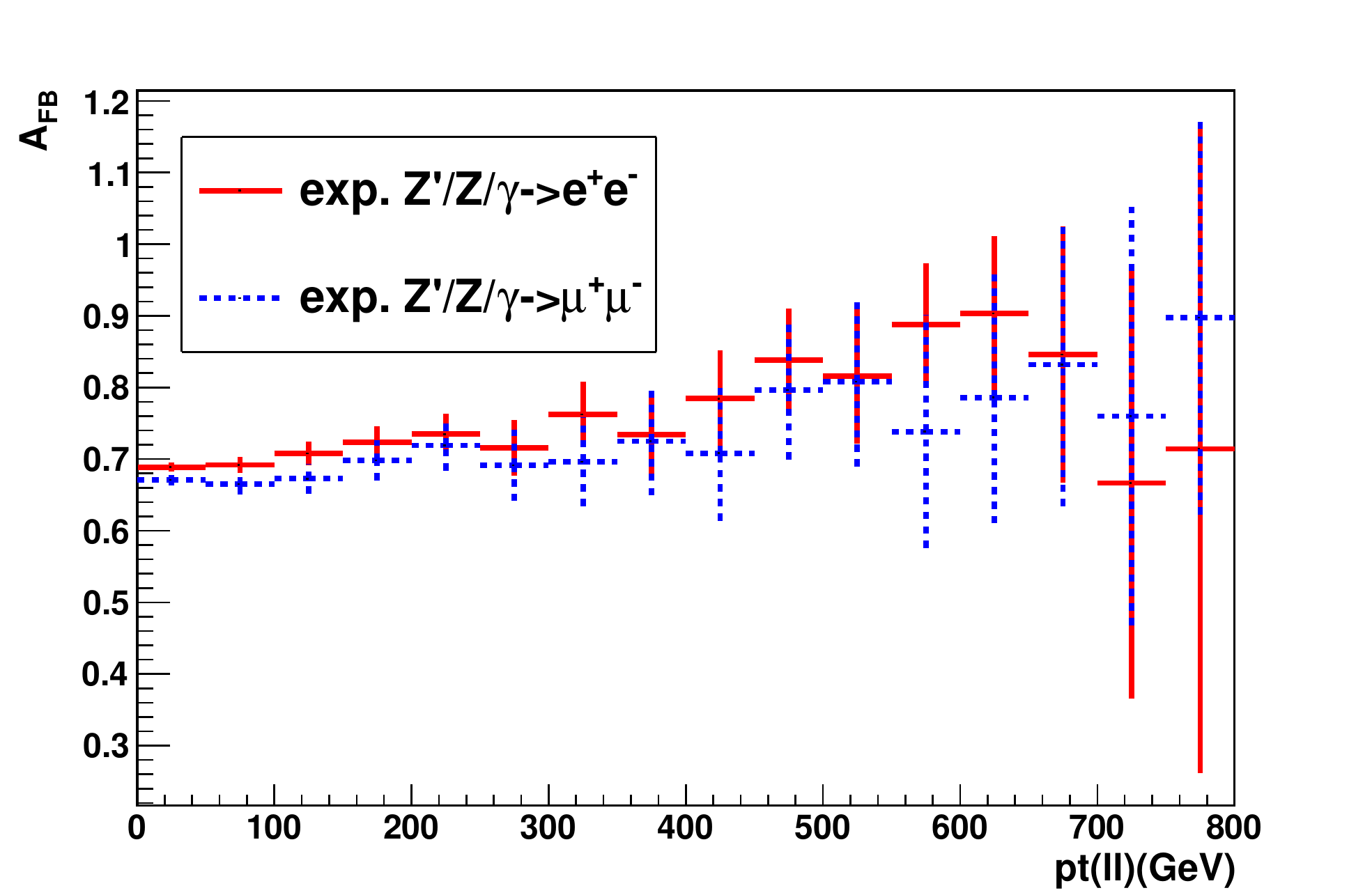}
    \caption{The forward backward asymmetry distribution as a function of the dilepton transverse momentum for $M_{Z^{\prime}} = 1.5$ TeV in the case of $\sqrt{s} = 14$ TeV with $500$ fb$^{-1}$ of data. The red lines represent the electron channel and the dotted blue lines denote the muon channel. The error bars are the statistical uncertainties normalized to $500$ fb$^{-1}$ of data.\protect\label{fig:AFB2}}}
\end{figure}

\section{Implications of $U(1)^{\prime}$ Symmetry for Sparticle Spectrum}
\label{sec:sparticle}

In addition to giving rise to realistic masses and mixing angles of the SM particles, including those of the neutrinos, the extra $U(1)^{\prime}$ symmetry can in general also dictates the mass spectrum of the sparticles, given the mediation mechanism of supersymmetry breaking.

\subsection{An Example with Anomaly Mediated SUSY Breaking}

The general soft SUSY breaking Lagrangian is given by,
\begin{equation}
\mathcal{L}_{soft} = -(m^2)_j^i \phi^i \phi^j - \biggl( \frac{1}{2} b^{ij} \phi^{i} \phi^{j} + \frac{1}{6} h^{ijk}\phi_i \phi_j \phi_k + \frac{1}{2} M_a \lambda_a \lambda_a + h.c. \biggr) \; ,
\end{equation}
where $M_a$ $(a=1,2,3)$ are the mass terms of the gaugino $\lambda_{a}$, $b^{ij}$ and $h^{ijk}$ are the bi-linear and  tri-linear terms, respectively, and $(m^2)^{i}_{j}$ are the scalar squared mass terms. 

Among various supersymmetry (SUSY) breaking mechanisms, Anomaly Mediated SUSY Breaking (AMSB)\cite{ref:SUSYAMSB} turns out to be an extremely predictive framework, in which the soft masses for the sparticles are generated by the conformal anomaly. As a result, {\it all} soft masses are determined entirely by the low energy dynamics ({\it i.e.} that of the MSSM) and one single parameter, $M_{\mbox{\tiny aux}}$, the F-term of some compensator chiral superfield. Explicitly, 
\begin{eqnarray}
\label{eqn:RGE1}
M_a = m_{3/2} \beta_{g_a}/g_a, \\
\label{eqn:RGE2}
h^{ijk} = -m_{3/2} \beta_Y^{ijk}, \\
\label{eqn:RGE3}
(m^2)_j^i = \frac{1}{2} m_{3/2}^2 \mu \frac{d}{d \mu} \gamma_j^i, \\
\label{eqn:RGE4}
b^{ij} = \kappa m_{3/2} \mu^{ij} - m_{3/2} \beta_{\mu}^{ij},
\end{eqnarray}
where $\gamma_j^i$ are the anomalous dimensions of the chiral superfields, $\mu^{ij}$ are the $\mu$ terms and $\beta_{g_a}$, $\beta_{Y}$ are the $\beta$-functions of the gauge and Yukawa couplings. With proper normalization, the F-term $M_{\mbox{\tiny aux}}$ is taken to be the gravitino mass, $m_{3/2}$. 
This is in stark contrast to the generic MSSM, where 124 parameters are present mostly to account for the soft SUSY breaking sector. 

\subsection{The $U(1)^{\prime}$ Symmetry and the Slepton Mass Problem}

The high predictivity also leads to a severe problem in AMSB models as generically the slepton masses are predicted to be tachyonic, because the electroweak gauge groups, $SU(2)_{L}$ and $U(1)_{Y}$, of the MSSM are not asymptotically free. Squarks do not suffer from the same problem as $SU(3)_{c}$ is asymptotically free.

The presence of an extra non-anomalous $U(1)^{\prime}$ provides a RG invariant solution to the tachyonic slepton mass problem\footnote{The $U(1)^{\prime}$ symmetry breaking scale in this case is close to the GUT scale. Consequently, the resulting $Z^{\prime}$ boson is not accessible by the currently collider experiments.}. In the presence of the $U(1)^{\prime}$, there are additional Fayet-Illiopolous (FI) D-term contributions to the scalar squared masses. Including the additional FI-D term contributions to the scalar masses, the new scalar squared masses can be written as,
\begin{eqnarray}
\begin{array}{lll}
\bar{m}_{Q}^2 = m_{Q}^2 + \zeta q_{Q_i} \delta_{j}^{i} \; , \quad  &
\bar{m}_{u^c}^2 = m_{u^c}^2 + \zeta q_{u_i} \delta_{j}^{i} \; , \quad &
\bar{m}_{d^c}^2 = m_{d^c}^2 + \zeta q_{d_i} \delta_{j}^{i} \; ,  \label{eqn:FIDMass1}\\
\bar{m}_{L}^2 = m_{L}^2 + \zeta q_{L_i} \delta_{j}^{i} \; , \quad &
\bar{m}_{e^c}^2 = m_{e^c}^2 + \zeta q_{e_i} \delta_{j}^{i} \; ,  & \label{eqn:FIDMass2}\\
\bar{m}_{H_u}^2 = m_{H_u}^2 + \zeta q_{H_u} \; , \quad &
\bar{m}_{H_d}^2 = m_{H_d}^2 + \zeta q_{H_d} \; . & 
\label{eqn:FIDMass3}
\end{array}
\end{eqnarray}
$m_{Q}^2$, $m_{u^c}^2$, etc denote the AMSB contributions to the scalar squared masses. The parameter $\zeta$ is the effective Fayet-Iliopoulos term setting the overall scale of the D-term contribution. 

Similar to the discussion in the previous sections, if the matter superfields have generation dependent $U(1)^{\prime}$ charges, then the $U(1)^{\prime}$ symmetry can play the role of a family symmetry giving rise to all SM fermion masses and mixing angles. The same set of charges also enter into the predictions for the sfermion masses, as dictated by Eq.~\ref{eqn:FIDMass1}. In addition, the anomaly cancellation conditions, which are required to ensure the RG invariance of the solutions, give very strong constraints on these charges.

\subsection{Sparticle Mass Spectrum and Sum Rules}

One characteristic feature of AMSB in the presence of D-term contributions is the existence of sum rules among squared masses of the sparticles. As the $U(1)^{\prime}$ symmetry in our model is generation-dependent and non-anomalous, the sum rules in this model are quite distinct from those found in other AMSB models with $U(1)^{\prime}$ symmetry\cite{ref:sumRuleAC}. The anomaly cancellation constraints lead to the D-term contributions among various fields to be cancelled automatically. Hence, the sum of the modified masses squared is still equal to the sum of mass square from the original AMSB contribution. The anomaly cancellation conditions $[SU(3)]^{2} U(1)^{\prime}$, $[SU(2)_{L}]^{2} U(1)^{\prime}$, $\left[U(1)_{Y}\right]^{2} U(1)^{\prime}$, 
give rise to the following RG invariant mass sum rules, and
\begin{eqnarray}
\label{eqn:sumSu3U1}
\sum_{i=1}^{3} (\bar{m}_{u_i^c}^2 + \bar{m}_{d_i^c}^2 + 2\bar{m}_{Q_i}^2) = \sum_{i=1}^{3} (m_{u_i^c}^2 + m_{d_i^c}^2 + 2m_{Q_i}^2)_{\mbox{\tiny AMSB}} \, ,  
\end{eqnarray}
\begin{eqnarray}
\label{eqn:sumSu2U1}
\sum_{i=1}^{3} (\bar{m}_{L_i}^2 + 3 \bar{m}_{Q_i}^2) & = & \sum_{i=1}^{3} (m_{L_i}^2 + 3m_{Q_i}^2)_{\mbox{\tiny AMSB}} \, , \\
\label{eqn:sumU1Y2U1}
\sum_{i=1}^{3} (\bar{m}_{u_i^c}^2 + \bar{m}_{e_i^c}^2 - 2\bar{m}_{Q_i}^2) & = & \sum_{i=1}^{3} (m_{u_i^c}^2 + m_{e_i^c}^2 - 2 m_{Q_i}^2)_{\mbox{\tiny AMSB}} 
\, ,
\end{eqnarray}
where terms on the right-hand side are the pure AMSB contributions, which are given in terms of $m_{3/2}^2$ and coefficients that are determined by the low energy dynamics (i.e., the gauge coupling constants and Yukawa coupling constants of MSSM). Similarly, sum rules within each generation i and j ($i, j = 1, 2, 3$) can be derived from the $U(1)^{\prime}$ gauge invariance\cite{ref:sumRuleGI}, 
\begin{eqnarray}
\label{eqn:quhu}
\bar{m}_{Q_i}^2 + \bar{m}_{u_j^c}^2 + \bar{m}_{H_u}^2 & = & (m_{Q_i}^2 + m_{u_j^c}^2 + m_{H_u}^2)_{\mbox{\tiny{AMSB}}} + (q_{Q_i} + q_{u_j} + q_{H_u}) \zeta \; \, , \\
\label{eqn:qdhd}
\bar{m}_{Q_i}^2 + \bar{m}_{d_j^c}^2 + \bar{m}_{H_d}^2 & = & (m_{Q_i}^2 + m_{d_j^c}^2 + m_{H_d}^2)_{\mbox{\tiny{AMSB}}} + (q_{Q_i} + q_{d_j} + q_{H_d}) \zeta \; \, , \\
\label{eqn:lehd}
\bar{m}_{L_i}^2 + \bar{m}_{e_j^c}^2 + \bar{m}_{H_d}^2 & = & (m_{L_i}^2 + m_{e_i^c}^2 + m_{H_d}^2)_{\mbox{\tiny{AMSB}}} +(q_{L_i} + q_{e_j} + q_{H_d}) \zeta \; \, .
\end{eqnarray}

Eqs. (\ref{eqn:sumSu3U1}-\ref{eqn:sumU1Y2U1}) then lead to the following sum rules for the physical masses,
\begin{eqnarray}
m_{\tilde{u}_L}^2 + m_{\tilde{u}_R}^2 + m_{\tilde{d}_L}^2 + m_{\tilde{d}_R}^2 + m_{\tilde{c}_L}^2 + m_{\tilde{c}_R}^2 + m_{\tilde{s}_L}^2 + m_{\tilde{s}_R}^2 + m_{\tilde{t}_1}^2 + m_{\tilde{t}_2}^2 + m_{\tilde{b}_1}^2 + m_{\tilde{b}_2}^2  \nonumber \\
= 2\sum_{i = 1}^{3} (2m_{\tilde{Q}_i}^2 + m_{\tilde{u}_i^c} + m_{\tilde{d}_i^c}) + 2 \sum_{i = 1}^{3} (m_{u_i}^2 + m_{d_i}^2) \, , \hspace{1in} \\
m_{\tilde{e}_L}^2 + m_{\tilde{e}_R}^2 + m_{\tilde{\mu}_L}^2 + m_{\tilde{\mu}_R}^2 + m_{\tilde{\tau}_1}^2 + m_{\tilde{\tau}_2}^2 + m_{\tilde{u}_L}^2 + m_{\tilde{u}_R}^2 + m_{\tilde{c}_L}^2 + m_{\tilde{c}_R}^2 + m_{\tilde{t}_1}^2 + m_{\tilde{t}_2}^2 \nonumber \\
= \sum_{i = 1}^{3} (m_{\tilde{L}_i}^2 + m_{\tilde{e}_i^c}^2 + m_{\tilde{Q}_i}^2 + m_{\tilde{u}_i^c}^2) + 2\sum_{i = 1}^{3} (m_{e_i}^2 + m_{u_i}^2) \, . \hspace{1in} 
\end{eqnarray}
These mass sum rules can be tested at the collider experiments.

In addition to various sum rules, another characteristic attribute is that the degeneracy of the sfermion masses among the first two generations is lifted. In the generation independent $U(1)^{\prime}$ senario, the first two generations of the sfermions in each sector have the same masses individually. However, in the generation dependent $U(1)^{\prime}$ model, their mass squared splittings are proportional to the $U(1)^{\prime}$ charge splitting, i.e., 
\begin{equation}
m_{\tilde{f}_2}^2 - m_{\tilde{f}_1}^2 = \zeta (q_{f_2} - q_{f_1}) \; ,\quad  
\end{equation}
which are non-zero and RG invariant. Therefore, by measuring the mass splittings, it is possible to distinguish various $U(1)^{\prime}$ models by identifying the charge splittings. 

\section{Conclusion}	
\label{sec:cond}

TeV scale mechanisms for neutrino mass generation afford the possibility of testing the mechanisms at low energy collider experiments. These models differ in their underlying symmetries as well as particle contents, leading to different signatures at the collider experiments and low energy lepton flavor violation searches. With the commissioning of the LHC and the SuperB Factory as well as various LFV searches including MEG and Mu2e, it may thus be possible to reveal the origin of neutrino mass generation, if it indeed takes place at the TeV scale.

\section*{Acknowledgments}

The work was supported, in part, by the National Science Foundation under Grant No. PHY-0709742 and PHY-0970173.

\end{document}